# Transfer pricing manipulation, tax penalty cost and the impact of foreign profit taxation

**Alex Augusto Timm Rathke**

*alex.rathke@usp.br*

*School of Economics, Business and Accounting at Ribeirão Preto - University of São Paulo*
*Address: Avenida dos Bandeirantes 3900, 14040-905 Ribeirão Preto, SP, Brazil.*

**Abstract**

This paper analyzes the optimal level of transfer pricing manipulation when the expected tax penalty is a function of the tax enforcement and the market price parameter. The arm's length principle implies the existence of a range of acceptable prices shaped by market, and firms can manipulate transfer prices more freely if market price range is wide, or if its delimitations are difficult to determine. Home taxation of foreign profits can reduce income shifting incentive, depending on the portion of repatriation for tax purposes. We find that the limited tax credit rule tends to be a less efficient measure, nonetheless it is the most widely adopted rule by countries, so to spark the perspective of more powerful approaches for taxation of foreign profits.

## 1. Introduction

Tax administrators from major economies unveil increasing concerns about international income shifting activities of multinational enterprises (MNE), as a strategy to reduce their global taxation. The strategy consists in transferring profits from high-tax locations to low-tax locations, so to subject a greater fraction of total profits to lower tax rates. Evidences indicate that income shifting is a worldwide persistent practice in several economic sectors (Devereux & Maffini, 2007; Borkowski, 1996; Hines, 1996) and the manipulation of prices in intra-firm transactions is a disseminated scheme to achieve it (Clausing (2003) and Swenson (2001) provide substantial evidences). In response to this tax avoidance conduct, tax authorities impose the adoption of law-enforced specific conditions for prices of internal transactions, and implement procedures to prevent outflows of profits or to reduce its effect on domestic tax base.

In intra-firm transactions, the top requirement is the application of the arm's length principle for the determination of transfer prices, in order to reflect the prices settled in conditions of commercial and financial independence. Despite formal imposition of this requirement, its effectiveness highly depends on stringency of tax authorities, regarding both the means to confirm the application of the principle and the acceptance of proofs prepared by MNE to sustain pricing choices (Lohse *et al*, 2012). In particular, arm's length condition involves significant complexity to tax auditors because it is a market-based criterion, so they necessarily rely on observable pricing information in external transactions for the assessment of transfer pricing appropriateness.

A second relevant mechanism against income shifting is the taxation of foreign profits by MNE's residence country: this is included in the controlled foreign company (CFC) rules. In this system, home country demands the taxation of profits earned abroad when they are transferred to the headquarter division, and establishes a type of compensation structure to avoid multiple taxation on the same profit. CFC rules typically carry clauses and provisions to regulate maintenance of profits in host country, aiming to inhibit abuse of repatriation deferral, and it commonly results in compulsory repatriation of a portion of foreign profits for tax purposes. In this case, even if taxable profits are transferred to distinct jurisdictions, MNE's home country is able to recover tax revenues through repatriation. This is a binding case if MNE's home country has a higher tax rate and suffers from outflow of profits, but it poses as an inefficient measure if profits are shifted away from host countries with high tax rates[1]. Hence, home taxation of foreign profits can influence transfer pricing manipulation, and the impact of distinct CFC rules are specially important for the optimal transfer price.

This study proceeds with further analysis of the transfer pricing manipulation when MNE is subjected to tax penalization if mispricing is detected. Tax penalty refers to an additional cost with some probability for its incurrence, and this expectation is associated with the level of tax enforcement exercised in the assessment of the transfer prices. We propose the segregation of endogenous and exogenous components of penalization probability, with respect to MNE, so to segregate the influence of countries' tax enforcement and of MNE's transfer pricing choice on the expected tax penalty cost. We then present a specification of the endogenous probability that aims to approximate the realistic relation between transfer prices and the arm's length parameter in a plain and comprehensible configuration[2]. The resulting model allows to demonstrate that the tax enforcement, denoting the exogenous probability and the tax fine, is negatively related with the optimal level of transfer pricing manipulation. As a main analysis from the model, application of arm's length principle suggests that the transfer price that stirs the tax penalty is not a discrete figure, but it depends on the range of acceptable values that are delineated by market participants. The delimitations of the accepted price range may be more or less observable, creating a critical difficulty for the evaluation of intra-firm prices by tax authorities. The model is applied to examine the effect of different rules of foreign profit taxation. In general, analysis indicates that income shifting incentive is not completely neutralized if there is a chance for MNE to manage the repatriation of foreign profits, and the limited tax credit rule, which is the one most widely adopted by countries (Hines, 2008; Schjelderup, 1999), appears to be not fully efficient in discouraging transfer pricing manipulation. The proportional tax credit rule may act as a more efficient alternative for discouraging income shifting and recovering tax revenues for home countries.

---

[1] Location of MNE's headquarter plays a relevant role in income shifting practices, for it can attract more profits even if it is not the most tax advantageous option (Dischinger *et al*, 2013).

[2] Existing studies limit the breakdown of the tax penalty function into factors scaling for the concealment cost, and fewer ones explicitly consider the penalization probability. To the best of our knowledge, the influence of arm's length range on the analysis of the expected tax penalty cost is still lacking.

Our analysis have potential implications for the current international discussion on the improvement of anti-shifting mechanisms and the issues regarding the application of arm's length principle on tax assessment of contemporaneous business. The Organization for the Economic Co-operation and Development (OECD) recognizes that MNE have been able to misapply current transfer pricing rules in order to allocate taxable profits to low-tax jurisdictions. Specifically, the prevailing arm's length system exhibits limitations when employed to the valuation of transactions with higher risks, and new developments on the application of arm's length principle are required to solve these obstacles (OECD, 2013). OECD highlights the impact of efficient CFC rules in context of transfer pricing manipulation, claiming that the strengthening of these rules can neutralize undue benefits from long-term deferral and reduce income shifting incentive. Although other alternatives for international income allocation have been proposed, OECD sustains its position to maintain the arm's length principle as the fundamental basis of the guidelines, to resolve the flaws in existing transfer pricing rules, and to enhance the design of existing CFC rules. In this line, our analysis aims to refine the understanding of the influence of the arm's length principle on transfer pricing choice, and to expose the outcomes of distinct rules of foreign profit taxation.

This study also contributes to recent researches that emphasize the influence of tax enforcement and tax regulations against income shifting. Literature sustains that income shifting implies concealment costs (Buettner *et al*, 2011; Haufler & Schjelderup, 2000), which reflect both the risk of supplementary tax and penalty payments for price manipulation, and MNE's extra efforts to justify the economic motivations of intra-firm transactions and to eventually hide opportunistic behavior in transfer pricing choice[3]. While the latter type is associated with transaction costs and harbor expenses incurred by the MNE, the effect of transfer pricing manipulation tends to have a chief role within the expected tax penalty cost of the former type[4], which is in the spot of this study. The probability of tax penalization is closely associated with the level of tax enforcement and anti-shifting legislation in each country, and recent studies find that these measures are efficient in preventing income shifting. Beer and Loeprick (2015), Beuselink *et al* (2014) and Lohse and Riedel (2013) obtain significant findings demonstrating that rigid transfer pricing rules reduce income shifting, and specific tax fines can exert additional enforcement towards tax compliance[5]. Results indicate, however, that transfer pricing regulation is not effective to control transmissions of intangibles (Beer & Loeprick, 2015), and this may be linked with the issues attributed to the application of current arm's length system to valuation of these transactions. Regarding CFC rules, Markle (2011) find that firms subjected to tax exemption of foreign profits shift more income from high-tax to low-tax locations, in comparison to firms that are imposed to residence-based taxation, and Egger and Wamser (2015) indicate that CFC rules bear ponderous costs for firms and restrain real investment abroad, thus limiting the opportunities for income shifting.

---

[3] Hines and Rice (1994) consider extended concealment costs, related with inefficiencies in internal transactions and additional investments to make pricing choices seem plausible. These costs appears to be particularly relevant if income shifting involves simulations and multiple-stage transactions, mainly by manipulation of transacted quantities.

[4] Evidence indicates that stronger signs of transfer pricing manipulation are associated with both higher level of audit activity and with higher returns generated from tax audits. See Alm (2012) for a survey on controlling tax evasion.

[5] Results indicate that, depending on the inferential model, tight transfer pricing rules can reduce income shifting by 50%.

The outline of this study is organized as follows: Section 2 structures the basic model for the benchmark case of tax exemption of foreign profits[6], including the expected tax penalty cost. We construct the specification of the endogenous component of tax penalty probability and derive the optimal level of transfer pricing manipulation affected by tax enforcement and arm's length parameter. In Sections 3-6, we analyze the effect of three rules of foreign profit taxation on the optimal transfer pricing manipulation. Section 7 extends the analysis for the impact of *ad valorem* tariffs on imports, and Section 8 concludes.

## 2. The model

Consider a MNE with two divisions, the parent company located in Country 1 and a wholly owned subsidiary located in Country 2 ($i = 1,2$). Both divisions produce outputs $x_i$, with costs $C_i(x_i)$, bringing revenues $R_i(s_i)$. Parent firm also exports part of its output ($m$) to subsidiary in Country 2, charging a transfer price $p$. The pretax profits of both divisions are

$$\pi_1 = R_1(s_1) - C_1(s_1 + m) + pm \quad \text{and} \quad \pi_2 = R_2(s_2) - C_2(s_2 - m) - pm$$

For an income tax rate $t_i$ in each country, application of source principle on taxation of foreign profits leads to MNE's global net profit

$$\Pi = (1-t_1)\pi_1 + (1-t_2)\pi_2 \quad (1)$$

Income shifting incentive arises when $t_1 \neq t_2$, so MNE maximizes $\Pi$ when greater portions of profits are transferred from high-tax to low-tax country, via manipulation of $p$. In Equation (1), we see that, if $t_2 > t_1$, $\Pi_p > 0$[7], and we call the high transfer price – *HTP* case; if $t_2 < t_1$, $\Pi_p < 0$, and we call the low transfer price – *LTP* case[8].

In order to discourage transfer pricing manipulation, both countries settle their own non-negligible and non-deductible tax penalty $Z_i$, charged if the difference between $p$ and a parameter price (assume an arm's length price $\bar{p}$) causes profits to be shifted away. Based on the determinants sustained in literature (Lohse & Riedel, 2013; Haufler & Schjelderup, 2000; Kant, 1988), we assume the incurrence of $Z_i$ depends on two main factors: the extent of transfer pricing manipulation and the level of country's tax enforcement.

First, probability of imposition of penalty $Z_i$ depends on the magnitude of the difference $p - \bar{p}$: a greater deviation from parameter $\bar{p}$ is more likely to be interpreted as tax evasion by fiscal authorities, and presents less defense chances in court. This factor is referred as probability $\alpha$. Second, stricter transfer pricing regulation and more rigorous tax audits represent higher probability of penalization. The influence of countries' enforcement activity is referred as probability $\varphi_i$. Thus, there are four outcomes derived from probabilities $\alpha$ and $\varphi_i$ (Table 1). When MNE manipulates $p$, there is some probability $\varphi_i$ that the harmed country $i$ detects mispricing, so $Z_i$ is inflicted. Indeed, there is probability $1 - \varphi_i$ that transfer pricing manipulation is unnoticed, implying no penalization. Further, when MNE does not manipulate $p$, complemental probability $1 - \alpha$ results in no tax penalty, regardless the level of tax enforcement.

---

[6] This is the benchmark case, with no taxation of foreign profits. Source-based taxation stands for the case of tax exemption of foreign profits.

[7] (Double) subscripts denote first (second) derivatives with respect to indicated variables.

[8] These simple conditions suffer modifications when additional factors are introduced in the analysis. We examine the more complex cases of residence principle taxation and the impact of *ad valorem* tariffs on imports in subsequent sections of this study, where *HTP* and *LTP* cases modify.

*Table 1: Outcomes from probabilities α and $\varphi_i$*

|  |  | Country's penalty imposition | |
|---|---|---|---|
|  |  | $\varphi_i$ | $1 - \varphi_i$ |
| Firm's penalty exposure | $\alpha$ | $\alpha\varphi_i Z_i$ | $\alpha(1 - \varphi_i)0$ |
|  | $1 - \alpha$ | $(1 - \alpha)\varphi_i 0$ | $(1 - \alpha)(1 - \varphi_i)0$ |

For a penalty $Z_i(m) = z_i m$, $z_i > 0$, unitary on the volume of intra-firm trade, the expected tax penalty due to transfer pricing manipulation is

$$E(Z_i) = \alpha \varphi_i z_i m \quad (2)$$

For the analysis, we relate to the probability components $\alpha$ and $\varphi_i$ as the following:

*Definition 1:* The expected tax penalty of transfer pricing manipulation as in Equation (2) is composed by two probability components: component $\alpha$ is the endogenous probability and component $\varphi_i$ is the exogenous probability, with respect to MNE.

In effect, probability $\alpha$ is endogenous because it reflects the impact of managements' transfer pricing decisions over the expected tax penalty, i.e. MNE is able to influence $E(Z_i)$ since changes in $p$ cause changes in $\alpha$. On the other hand, probability $\varphi_i$ is exogenous because it characterizes governments' exercise to detect and punish income shifting, which suffers no influence from MNE's decisions. Although tax rules require the application of arm's length principle, countries' actions related with enforcing audit procedures and strengthening fiscal interpretation of what is an acceptable transfer price cannot themselves change the market-based parameter $\bar{p}$; so, tax authorities are either not able to influence $\alpha$. Thus, because of these properties, components $\alpha$ and $\varphi_i$ are assumed to be independent[9].

MNE faces additional cost of the expected penalty and focuses in maximizing the objective function $\phi = \Pi - E(Z_i)$[10]. For a tax enforcement $\varphi_i > 0$, manipulation of $p$ implies $E(Z_i)_p > 0$, while $\phi_{E(Zi)} < 0$. Therefore, profit maximization occurs when MNE achieves equilibrium between reducing global income tax and increasing expected tax penalty.

*2.1 Specification of endogenous probability α and optimal level of transfer pricing manipulation*

To design a reasonable specification for $\varphi_i$ is a rather complex task, because the estimation of degrees of tax enforcement relies upon government-level factors that are difficult to observe and measure[11]. Besides the existence of specific tax rules, exogenous probability $\varphi_i$ depends on governments' mechanisms to guarantee proper application of these rules, which are associated mainly with audit strategies, the effectiveness of fiscal monitoring and the characteristics of legal system.

---

[9] The assumption reflects the economic concept of price-taker for governments. This is consistent with the arm's length system in OECD guidance.
[10] $i = 2$ for *HTP* case, and $i = 1$ for *LTP* case.

Diversely, specification of $\alpha$ appears to be a more straightforward matter. Literature on tax regulation assumes the standard model as a twice-differentiable function $f(p - \bar{p})$, satisfying $f(p = \bar{p}) = 0$, sign $f_p$ = sign $(p - \bar{p})$ and $f_{pp} > 0$ (Buettner *et al*, 2011; Haufler & Schjelderup, 2000; Kant, 1988); i.e. in our model, standard conditions are internalized in $\alpha$. With respect to parameter price $\bar{p}$, independent negotiations in open market usually generate a range of market prices, all of which represent the prices obtained in transactions under arm's length conditions (OECD, 2010). It comprises the accepted variance of $\bar{p}$ due to market forces, so any $p$ within this range is considered an appropriate market price in some degree. In this study, we follow the concept of arm's length conditions stated in OECD guidance[12].

Assume a market price range $p^{min} < \bar{p} < p^{max}$, where $p^{min} > 0$, so $\bar{p}$ is the central parameter[13]. $p^{min}$ and $p^{max}$ are bottom and top limits of market price range, respectively, and are outlined by (presumably tacit) consensus of market players. In this way, $p^{max}$ ($p^{min}$) is the limiting price which triggers tax penalty with certainty for *HTP* (*LTP*) case (Kant, 1988)[14], i.e. $\alpha$ increases as $p$ gets closer to $p^c$ ($c = min, max$). Inversely, there is no tax penalty if $p$ is settled at or below (above) $\bar{p}$ in *HTP* (*LTP*) case. Therefore, endogenous probability $\alpha$ can be specified as

$$\alpha = \frac{(p - \bar{p})^r}{(p^c - \bar{p})^r} \text{ , for } c = min, max \quad (3)$$

where $r$ is the curve's slope. From Equation (3), we set:

- for *HTP*, $c = max$; $\begin{cases} \bar{p} < p < p^{max} & \to \quad 0 < \alpha < 1 \\ \bar{p} \geq p & \to \quad \alpha = 0 \\ p \geq p^{max} & \to \quad \alpha = 1 \end{cases}$

- for *LTP*, $c = min$; $\begin{cases} p^{min} < p < \bar{p} & \to \quad 0 < \alpha < 1 \\ \bar{p} \leq p & \to \quad \alpha = 0 \\ p \leq p^{min} & \to \quad \alpha = 1 \end{cases}$

---

[11] Some studies create measures of legal enforcement and stringency based on rules characteristics, e.g. Lohse *et al.*, 2012.

[12] Note that the arm's length principle does not require an existing active market for the objects of the transaction. It states that the arm's length price is the price binding the conditions an open market would demand, in normal operation. While there may be genuine difficulty in determining the appropriate transfer price in absence of market forces, the principle has become sufficiently familiar to allow a common understanding among governments and firms (OECD, 2010), so they are still capable to find a consensus of an accepted arm's length range. The range's delimitations may be more or less blurred, depending on the observable information used to reach consensus.

[13] $\bar{p}$ can be interpreted as the closest figure to some ideally accurate arm's length price. Here, it can be assigned as the mean price.

[14] *HTP* case refers to $p - \bar{p} > 0$; *LTP* case refers to $p - \bar{p} < 0$.

Equation (3) satisfies standard conditions[15] for all $r > 1$. It models the endogenous probability $\alpha$ by mirroring the relation between transfer price $p$ and price parameters shaped by market, thus $\alpha$ depends on deviation of $p$ from $\bar{p}$, with respect to market parameter $p^c - \bar{p}$.

The optimal level of transfer pricing manipulation can be analyzed differentiating the objective function $\phi$ with respect to $p$, using $\alpha$ as specified in Equation (3). Assuming $r = 2$ for simplification, we obtain

$$\phi_p = (t_2 - t_1)m - 2\frac{p - \bar{p}}{(p^c - \bar{p})^2}\varphi_i z_i m = 0 \quad (4)$$

Clearly, the level of transfer pricing manipulation is expressed by the term $p - \bar{p}$. Solving Equation (4) for $p - \bar{p}$ leads to

$$p - \bar{p} = \frac{t_2 - t_1}{\varphi_i z_i}\frac{(p^c - \bar{p})^2}{2} \quad (5)$$

Equation (5) provides interesting understanding on factors influencing firms' motivations to shift profits among countries when they are subjected to possible tax penalization (details in **Appendix A**). Initially, tax differential $t_2 - t_1$ preserves the position of prime incentive for income shifting, as raised by seminal work of Horst (1971). The magnitude of tax differential essentially computes for the extent of transfer pricing manipulation, and its sign determines the direction of transfers.

Moreover, denominator $\varphi_i z_i$ on right hand of Equation (5) accounts for conjoint factors of tax penalization in case that mispricing is detected. Increase of either exogenous probability $\varphi_i$ or tax penalty $z_i$ reduces the weight of tax differential, so the optimal level of transfer pricing manipulation is reduced. Countries are able to enforce tax actions in order to discourage income shifting. Larger fines produce larger potential costs when misbehavior is identified, and MNE becomes exposed to bigger losses. Besides that, governments can increase tax enforcement through specific tax rules, while auditors can toughen tax inspections and improve efficiency of tax controls, leading to a greater probability of imposition of tax penalty. Even if unitary penalty $z_i$ is not changed, an increase of exogenous probability $\varphi_i$ causes optimal $p$ to get closer to $\bar{p}$, i.e. increase in $\varphi_i$ weaken the impact of tax differential, so MNE is forced to reduce the distance between $p$ and $\bar{p}$ in order to restore equilibrium.

Equation (5) also shows that the optimal level of transfer pricing manipulation is influenced by $p^c - \bar{p}$. A wide range of market prices provides a whole set of alternatives for MNE to choose a transfer price in accordance with arm's length criteria. Formally, expanding the range between $p^c$ and $\bar{p}$ causes an increase in the level of transfer pricing manipulation because it allows MNE to enlarge the distance between $p$ and $\bar{p}$ and still be tax compliant. Any $p$ within price parameter $p^c - \bar{p}$ satisfies arm's length condition required by tax authorities, so MNE is able to justify its choice against questionings and prosecutions whether $p$ does not hit $p^c$, and this can be specially beneficial for MNE if $p^c$ is hard to delimit.

In what follows, we analyze transfer pricing manipulation when foreign profits are subjected to residence-based taxation rules in home country. First, we consider the case when home country allows tax credits for foreign taxes paid abroad, those credits being limited by the proportion of the amount of foreign profits repatriated by MNE: this is the proportional tax credit rule. The second case is when tax credits are allowed in home country, but are limited to the home taxation of foreign profits: this is the limited tax credit rule. In sequence, we analyze the third case when foreign taxes are treated as deductible costs in home country: this is the foreign tax deduction rule. Finally, a summary comparison is shown, and the impact of *ad valorem* tariffs on imports is pondered.

---

[15] Equation (3) applies for the instance $f_{pp} < 0$, e.g. $0 < r < 1$.

**3. Proportional tax credit rule and transfer pricing manipulation**

Assume Country 1 demands domestic taxation of foreign profits at the moment they are repatriated, and applies proportional tax credit rule on recoverable foreign taxes. Assume also that MNE is allowed to defer repatriation of foreign profits, and some amount of profits earned by subsidiary may remain[16] in Country 2. Let the rate of repatriation be $b$, where $0 \leq b \leq 1$. Given $t_1 \neq t_2$, MNE has incentive to manage $b$ in order to reduce global taxation, depending on which tax rate is higher.

For the *HTP* case, $t_2 > t_1$ implies that tax credit is equal to home taxation of foreign profits $t_1 b \pi_2$, because MNE is not allowed to use taxes paid in Country 2 to offset income tax[17] in Country 1. This ultimate restriction gives

$$\Pi = (1-t_1)\pi_1 + (1-t_2)\pi_2 - t_1 b \pi_2 + t_1 b \pi_2 = (1-t_1)\pi_1 + (1-t_2)\pi_2 \quad (6)$$

so global net income and the level of transfer pricing manipulation are not influenced by $b$ (Schjelderup, 1999; Kant, 1988).

For the *LTP* case, global net profits are

$$\Pi = (1-t_1)\pi_1 + (1-t_2)\pi_2 - t_1 b \pi_2 + t_2 b \pi_2 \quad (7)$$

Substituting Equation (7) in objective function $\phi$ and proceeding with same steps up to Equation (5) result in

$$p - \overline{p} = \frac{t_2(1-b) - t_1(1-b)}{\varphi_i z_i} \frac{(p^c - \overline{p})^2}{2} = \frac{(t_2 - t_1)(1-b)}{\varphi_i z_i} \frac{(p^c - \overline{p})^2}{2} \quad (8)$$

Comparing Equation (8) with Equation (5), we find that both equations present the same characteristics regarding the influence of tax enforcement and of market price parameter over the income shifting incentive. We find in Equation (8), though, that $b$ is negatively related with the level of transfer pricing manipulation, so that the repatriation of foreign profits reduces mispricing incentive by proportion $b$. We can state:

*Proposition 1:* For *LTP* case, proportional tax credit rule reduces the optimal level of transfer pricing manipulation by the proportion of the repatriation rate $b$.

*Proof:* See **Appendix B**. □

---

[16] A number of countries allows the deferral of repatriation of foreign profits, e.g. United States (Internal Revenue Code – IRC, Subpart F, §§951-965), Germany (*Außensteuergesetz – AStG*, §§7-14), and United Kingdom (Taxation (International and Other Provisions) Act 2010 – TIOPA, Part 9A).
[17] Any $b > 0$ makes the limit $t_1 b \pi_2 \geq t_2 b \pi_2$ is inconsistent with *HTP* case.

If foreign profits are retained in Country 2 they are taxed only by rate $t_2$, so MNE has incentive to choose $b = 0$. On the opposite stand, home country is inclined to require the repatriation of greater portion of foreign profits in order to counteract the impact of transfer pricing manipulation[18], i.e. mandatory repatriation brings shifted profits back for taxation at home country.

## 4. Limited tax credit rule and transfer pricing manipulation

Assume the circumstances stated in previous section, but now home country applies the limited tax credit rule. Here again, MNE has incentive to manage $b$ whether $t_1 \neq t_2$. MNE can use full tax credits, but faces the limitation $t_1 b \pi_2 \geq t_2 q \pi_2$, where $q$ is the rate of tax credit granted by Country 1, so $0 \leq q \leq 1$.

For both *HTP* and *LTP* cases, global net profits are

$$\Pi = (1-t_1)\pi_1 + (1-t_2)\pi_2 - t_1 b \pi_2 + t_2 q \pi_2 \quad (9)$$

formally subjected to inequality constraint $t_1 b \pi_2 \geq t_2 q \pi_2$. Solving $\phi_p$ for $p - \bar{p}$, with net profits in Equation (9), we get

$$p - \bar{p} = \frac{t_2(1-q) - t_1(1-b)}{\varphi_i z_i} \frac{(p^c - \bar{p})^2}{2} \quad (10)$$

for which previous properties regarding tax enforcement and market price parameter remain. Initially, if $q = b$, Equation (10) becomes identical to Equation (8), thus MNE has incentive to set a higher $q$, so greater tax credits are recovered. Indeed, MNE is able to offset home taxation of foreign profits if there is ability to manage $b$ within $0 \leq b \leq t_2/t_1$, and additional taxation is incurred when $b > t_2/t_1$ occurs[19]. We arrive at the following:

*Proposition 2:* Limited tax credit rule reduces the optimal level of transfer pricing manipulation if $b > t_2/t_1$.

*Proof:* See **Appendix C**. □

From *Proposition 2*, we can state:

*Corollary 1:* For $t_2 > t_1$, limited tax credit rule implies $q < b$, hence $t_1 b \pi_2$ is completely offset by $t_2 q \pi_2$.

Here, *Corollary 1* refers to *HTP* case. When maximizing condition $b = q(t_2/t_1)$ is satisfied, MNE neutralizes the effect of repatriation, and optimal level of transfer pricing manipulation assumes structure of Equation (5). It indicates that, if governments allow MNE to defer repatriation of a fraction of foreign profits, limited tax credit rule is not fully efficient in discouraging income shifting. The trigger for *LTP* case is when the rate of mandatory repatriation is high enough so taxation of $\pi_2$ by home country becomes greater than income tax in host country.

---

[18] If $b = 1$, Equation (8) becomes $p - \bar{p} = 0$. Thus, full repatriation neutralizes the incentive to manipulate transfer prices.

[19] Maximization condition $b = q(t_2/t_1)$, for limiting situation with maximum tax credit rate, $q = 1$.

# 5. Foreign tax deduction rule and transfer pricing manipulation

Assume the conditions from previous sections, with the difference that home country imposes the foreign tax deduction rule on repatriated foreign profits. MNE's global net income takes the form

$$\Pi = (1-t_1)\pi_1 + (1-t_2)\pi_2 - t_1 b(\pi_2 - t_2\pi_2) \quad (11)$$

The optimal level of transfer pricing manipulation becomes

$$p - \bar{p} = \frac{t_2(1-t_1 b) - t_1(1-b)}{\varphi_i z_i} \frac{(p^c - \bar{p})^2}{2} \quad (12)$$

Impacts of tax enforcement and of market price parameter remain. Foreign tax deduction rule brings a novel instance regarding maximization of global net income, because it can generate an overall taxation greater than $t_1(\pi_1 + \pi_2)$, e.g. full repatriation implies $\Pi = (1 - t_1)(\pi_1 + \pi_2) - t_2(\pi_2 - t_1\pi_2)$, so Equation (12) renders $p - \bar{p} \neq 0$ if $b = 1$. Even if $t_2 = t_1$, MNE still has incentive to manipulate transfer prices, depending on the magnitude of repatriation rate. In fact, we observe that $(p - \bar{p})_b > 0$ because an increase in $b$ causes optimal $p$ to increase (decrease in $b$ causes the opposite). We can state:

*Proposition 3:* In foreign tax deduction rule, optimal transfer price $p$ is positively related with repatriation rate $b$.

*Proof:* See **Appendix D**. □

Therefore, repatriation of foreign profits is capable of discouraging income shifting only in *LTP* case, because $p - \bar{p}$ is already positive in *HTP* case. In detail, *Proposition 3* derives:

*Corollary 2:* For $t_2 < t_1$, foreign tax deduction rule neutralizes the optimal level of transfer pricing manipulation (optimal $p - \bar{p}$ is zero) if $b = (t_1 - t_2)/[t_1(1 - t_2)]$.

MNE can face the particular situation where optimal $p - \bar{p}$ is positive (*HTP* case) while $t_2 < t_1$, i.e. for a sufficiently large $b$, higher $p$ can increase global net income, although home country levies the superior tax rate. In effect, MNE incurs in additional taxation $t_1 b\pi_2$ but recovers only $t_1 t_2 b\pi_2$; thus, total tax burden on foreign profits can be greater than one on domestic profits, and MNE would be in better position if both divisions were located in Country 1. Ultimately, maximization of global net income occurs when MNE has the alternative to retain $\pi_2$ completely in host country.

# 6. Summary comparison of foreign profit taxation rules

Comparing Equations (8), (10) and (12), we see that their differences rely on the recoverable portion of taxation in host country. For *LTP* case binding all three rules, comparison demonstrates that

$$\left(1-b\frac{t_1}{t_2}\right) < (1-b) < (1-t_1 b) \quad (13)$$

The term $1 - b(t_1/t_2)$ on the first left is the $t_2$-offset impact for the limited tax credit rule: it causes the smallest change in the optimal level of transfer pricing manipulation, specially when tax differential is large, because this rule allows the recovery of a higher credit based on taxes paid in Country 2. On the opposite right of Equation (13), $1 - t_1 b$ is the effect of the foreign tax deduction rule: repatriation under this rule produces a greater effect on optimal $p - \bar{p}$. This is due to the characteristics of double taxation arising from limiting compensation of foreign taxation up to $t_1 t_2 b \pi_2$. The intermediate term $1 - b$ comes from the proportional tax credit rule: it illustrates the outcome stated in *Proposition 1*. Equation (13) demonstrates in a clearer way that the influence of the proportional tax credit rule on income shifting incentive does not depend on tax rates of each country. This can be an attractive aspect for tax administrators, since the rule's applicability becomes plain for all locations, and the expected consequences on income shifting might become reasonably simpler to foresee.

## 7. Impact of tariffs on imports[20]

While home country may use the residence principle on taxation of foreign profits as an attempt to recover evaded tax revenues, host country has the alternative to impose *ad valorem* tariffs on imports ($\tau$), so the additional taxation on intra-firm transactions may discourage income shifting. As a consequence, $t_2 = t_1$ is not sufficient to annul optimal $p - \bar{p}$. See that

$$\tau = \frac{t_2 - t_1}{1 - t_2} \quad (14)$$

is the regular requirement to thoroughly cancel the incentive to manipulate transfer prices (Eden, 1983; Itagaki, 1982; Samuelson, 1982). It stands when home country applies the tax exemption rule on foreign profits[21]: if relative tax differential $(t_2 - t_1)/(1 - t_2)$ is higher than tariff $\tau$, maximization occurs at *HTP* case; if $\tau$ dominates, *LTP* is the maximization case.

When home country applies the proportional tax credit rule, imposition of $\tau$ shortens the impact of repatriation rate $b$. Mispricing incentive still reduces, but it occurs by less than proportion $b$ itself, so *Proposition 1* is dropped.

Regarding the limited tax credit rule, tariff $\tau$ affects both sides of restriction $t_1 b \pi_2 \geq t_2 q \pi_2$. It shows that MNE maximizes global net income when $b = q(t_2/t_1)$, which corresponds to the same conclusion obtained in the analysis of Equation (10). Thence, *Proposition 2* and *Corollary 1* hold in presence of tariff on imports.

In analysis of the foreign tax deduction rule, our conclusion is the same as one previously presented: we have $(p - \bar{p})_b > 0$ following from $p_b > 0$, so *Proposition 3* still applies. In addition, we find that, if $\tau$ is high enough, *LTP* case can become the maximization solution (even when $t_2 > t_1$). Tariff $\tau$ modifies necessary condition of $b$ to neutralize income shifting incentive, and it turns *Corollary 2* consistent for both *HTP* and *LTP* cases. *Corollary 2* is extended to account for the effect of $\tau$:

---

[20] Detailed analysis for this section in **Appendix E**.
[21] Condition in Equation (14) is altered if home country applies residence-based taxation on foreign profits.

*Corollary 2 – extended:* Foreign tax deduction rule neutralizes the optimal level of transfer pricing manipulation (optimal $p - \bar{p}$ is zero) if $b = [t_1 - t_2(1 + \tau) + \tau]/[t_1(1 + \tau)(1 - t_2)]$.

Comparing the impact of all three residence-based rules for taxation of $\pi_2$, we obtain the same inequality expressed in Equation (13)[22]. In general, MNE's responses to the rules of foreign profit taxation are essentially the same as for the absence of $\tau$, but in different magnitude. Taxation in Country 2 becomes a function of both $t_2$ and $\tau$, so it affects the amount of optimal level of transfer pricing manipulation and the weight of repatriation effect, still the result of changing from one rule to another is the same.

## 8. Conclusions

This paper examines the optimal level of transfer pricing manipulation, taking into account certain probability of tax penalization by the impaired country. The endogenous probability function proposed in this study is useful to derive a fair structure for the factors influencing income shifting, and we apply this model to analyze the effects of foreign profit taxation rules. The analysis indicates that home taxation of foreign profits can reduce mispricing incentive, but the effect depends specially on the amount of compulsory repatriation. In spite of the broad application, the limited tax credit rule tends to be a less efficient alternative, in comparison to the other residence-based rules analyzed. Interesting result is that foreign tax deduction rule may actually induce income shifting if mandatory repatriation rate is superlative. *Ad valorem* tariffs on imports reduce manipulation mainly at the expense of the firm, yet repatriation effect on recovered tax revenues by home country may be materially attenuated for large taxation in host country.

Our model provides a credible configuration for the influences of tax enforcement and market parameters on the transfer pricing choice, and can be applied to analyze these factors on other anti-shifting mechanisms. The model is consistent with the generally accepted role of tax enforcement to discourage manipulation. Besides, it incorporates the influence of the arm's length range, showing that this effect is likewise significant for the evaluation of transfer pricing appropriateness. The findings suggest that governments may face a hard time challenging transfer prices if market values presents high volatility or if market activity does not allow to accurately observe delimitations of the accepted price range, so firms may benefit from it. The influence of changes in the arm's length parameters is potentially relevant for studies on legislation efficiency and tax audits, and it appears to be a path for further investigation.

## References

Alm, J. (2012). Measuring, explaining, and controlling tax evasion: lessons from theory, experiments, and field studies. *International Tax and Public Finance* 19(1), pp. 54-77.

Amerighi, O. (2008). Transfer pricing and enforcement policy in oligopolistic markets. In *Foreign Direct Investment and the Multinational Enterprise* (pp. 117-154). Mass, Cambridge: MIT Press.

---

[22] Here, drop of *Proposition 1* does not affect the inequality, for the effect $\tau(1 - t_1 b)$ is the same in all three residence-based rules, so resulting in the same Equation (13).

Beer, S., & Loeprick, J. (2015). Profit shifting: drivers of transfer (mis)pricing and the potential countermeasures. *International Tax and Public Finance* 22(3), pp. 426-451.

Beuselink, C., Deloof, M., & Vanstraelen, A. (2014). Cross-jurisdictional income shifting and tax enforcement: evidence from public versus private multinationals. *Review of Accounting Studies*, pp. 1-37.

Borkowski, S. C. (1996). An analisys (meta- and otherwise) of multinational transfer pricing research). *The International Journal of Accounting* 31(1), pp. 39-53.

Clausing, K. A. (2003). Tax-motivated transfer pricing and US intrafirm trade prices. *Journal of Public Economics* 87, pp. 2207-2223.

Devereux, M., & Maffini, G. (2007). The impact of taxation on the location of capital, firms and profit: A survey of empirical evidence. *Oxford University Centre for Business Taxation WP 07/02*.

Dischinger, M., Knoll, B., & Riedel, N. (2013). The role of headquarters in multinational profit shifting strategies. *International Tax and Public Finance* 21(2), pp. 1-24.

Eden, L. (1983). Transfer pricing policies under tariff barriers. *Canadian Journal of Economics* 16(4), pp. 669-685.

Egger, P., & Wamser, G. (2015). The impact of controlled foreign company legislation on real investments abroad: a multi-dimensional regression discontinuity design. *Journal of Public Economics (in press)*.

Haufler, A., & Schjelderup, G. (2000). Corporate tax systems and cross country profit shifting. *Oxford Economic Papers* 52(2), pp. 306-325.

Hines Jr., J. R. (1996). Tax policy and the activities of multinational corporations. *National Bureau of Economic Research No. w5589*.

Hines Jr., J. R. (2008). Reconsidering the taxation of foreign income. *Tax Law Review* 62, pp. 269-298.

Hines Jr., J. R., & Rice, E. M. (1994). Fiscal paradise: Foreign tax havens and American business. *Quarterly Journal of Economics* 109, pp. 149-182.

Horst, T. (1971). The theory of the multinational firm: optimal behavior under different tariff and tax rates. *Journal of Political Economy*, pp. 1059-1072.

Itagaki, T. (1982). Systems of taxation of multinational firms under exchange risk. *Southern Economic Journal* 48(3), pp. 708-723.

Kant, C. (1988). Endogenous transfer pricing and the effects of uncertain regulation. *Journal of International Economics* 24(1), pp. 147-157.

Lohse, T., & Riedel, N. (2013). Do Transfer Prices Laws Limit International Income Shifting? Evidence from European Multinationals. *Oxford University Centre for Business Taxation Working Paper 13/07*.

Lohse, T., Riedel, N., & Spengel, C. (2012). The increasing importance of transfer pricing regulations - a worldwide overview. *Oxford Center for Business Taxation Working Paper WP 12/27*.

Markle, K. (2011). A comparison of the tax-motivated income shifting of multinationals in territorial and worldwide countries. *SSRN Paper no. 1764031*.

Organization for Economic Co-operation and Development. (2010). *OECD Transfer Pricing Guidelines for Multinational Enterprises and Tax Administrations*. Paris, França: OECD.

Organization for Economic Co-operation and Development. (2013). *Action Plan on Base Erosion and Profit Shifting*. Paris, França: OECD.

Samuelson, L. (1982). The multinational firm with arm's length transfer price limits. *Journal of International Economics* 13(3), pp. 365-374.

Schjelderup, G. (1999). Multinationals, intra-firm trade and the taxation of foreign-source income. *International Journal of the Economics of Business* 6(1), pp. 93-105.

Swenson, D. L. (2001). Tax reforms and evidence of transfer pricing. *National Tax Journal* 54(1), pp. 7-25.

**Appendix A: Analysis of optimal level of transfer pricing manipulation**

Assume the benchmark case with simplifications $G_i = \varphi_i z_i$ and $P = p^c - \bar{p}$ for tax enforcement[23] and price range variables, respectively. After the optimal level of transfer pricing manipulation derived from $\phi_p$, differential with respect to tax enforcement $G_i$ gives

$$(p - \bar{p})_{G_i} = -\frac{\dfrac{t_2 - t_1}{G_i^2} \dfrac{P^r}{r}}{(r-1)\left[\dfrac{t_2 - t_1}{G_i} \dfrac{P^r}{r}\right]^{\frac{r-2}{r-1}}} = -\frac{1}{(r-1)G_i}\left[\dfrac{t_2 - t_1}{G_i} \dfrac{P^r}{r}\right]^{\frac{1}{r-1}} < 0$$

which is negative for all $r > 1$, for both *HTP* and *LTP* cases[24]. If $\alpha = 1$ ($p \geq p^{max}$ in *HTP* case; $p \leq p^{min}$ in *LTP* case), $E(Z_i)$ is not a function of $p$. In any case, we find that $\phi_{Gi} = -E(Z_i)_{Gi} < 0$.

Differentiating $p - \bar{p}$ with respect to price range $P$ yields

$$(p - \bar{p})_P = \frac{r\dfrac{t_2 - t_1}{G_i} \dfrac{P^{r-1}}{r}}{(r-1)\left[\dfrac{t_2 - t_1}{G_i} \dfrac{P^r}{r}\right]^{\frac{r-2}{r-1}}} = \frac{r}{r-1}\left[\dfrac{t_2 - t_1}{G_i} \dfrac{P}{r}\right]^{\frac{1}{r-1}} > 0$$

which is positive for all $r > 1$, for both *HTP* and *LTP* cases. Even if changes in price range are due to changes in $\bar{p}$, we still have $(p - \bar{p})_P > 0$ because $(t_2 - t_1)P > 0$ (and semi-elasticity $r/(r-1) > 1$).

**Appendix B: Effect of repatriation under proportional tax credit rule for *LTP* case**

Differentiating Equation (8) with respect to $b$ provides

$$(p - \bar{p})_b = -\frac{t_2 - t_1}{\varphi_i z_i} \frac{(p^c - \bar{p})^2}{2}$$

which is matching with Equation (5), although is negative. It denotes that any increase (decrease) in $b$ causes a decrease (increases) in level of income shifting by the proportion $b$ itself, thus MNE maximizes global profits choosing the lowest possible repatriation rate. Derivative of Equation (7) with respect to $b$ is $\Pi_b = t_2\pi_2 - t_1\pi_2$. The repatriation rate that maximizes global net income is the one that equates taxation of $\pi_2$ in both countries; thus, maximization occurs when $b = 0$. Note that *LTP* case implies $p - \bar{p} < 0$. However, if $b > 0$ we have $(p - \bar{p})_b > 0$, which neutralizes the effect of transfer pricing manipulation. If full repatriation is mandatory, $b = 1$ and global net income from Equation (7) becomes $\Pi = (1 - t_1)(\pi_1 + \pi_2)$.

---

[23] In Equation (2), both exogenous probability $\varphi_i$ and tax penalty $z_i$ entail the same impact within the expected tax penalty costs.

[24] When n-degree polynomial derives two results (e.g. when $r$ is uneven), the solution is one consistent with the direction (sign) of income shifting for either *HTP* or *LTP* case.

**Appendix C: Effect of repatriation under limited tax credit rule**

MNE aims to maximize global profits, but faces legal limitation $t_1 b \pi_2 \geq t_2 q \pi_2$. For global net income in Equation (9), the Lagrangian of objective function is

$$L(p,b,q,\lambda) = (1-t_1)\pi_1 + (1-t_2)\pi_2 - t_1 b\pi_2 + t_2 q\pi_2 - \alpha\varphi_i z_i m + \lambda(t_1 b\pi_2 - t_2 q\pi_2)$$

The first order conditions are[25]

$$L_p = t_2(1-q)m - t_1(1-b)m - 2\frac{p-\bar{p}}{(p^c-\bar{p})^2}\varphi_i z_i m - \lambda(t_1 b - t_2 q)m = 0$$

$$L_b = -t_1 \pi_2 + \lambda t_1 = 0$$

$$L_q = t_2 \pi_2 - \lambda t_2 = 0$$

and the Kuhn-Tucker condition is $\lambda(t_1 b - t_2 q)m = 0$, where $t_1 bm \geq t_2 qm$, $\lambda \geq 0$. $L_p$ gives rise to Equation (10) restricted by $t_1 bm \geq t_2 qm$, and both $L_b$ and $L_q$ imply $\lambda = \pi_2$ (we get $\lambda > 0$). Therefore, Kuhn-Tucker condition is satisfied when $t_1 bm = t_2 qm$ and MNE maximizes global net income if $b = q(t_2/t_1)$. The restriction on tax credit causes $q$ vary according to variance of $b$.

For *HTP* case, the tax credit constraint is slack at the solution because it entails $q < b$, and MNE is able to completely offset foreign taxation in home country, even if full repatriation is mandatory. Global net income from Equation (9) becomes $\Pi = (1-t_1)\pi_1 + (1-t_2)\pi_2$ for any $b = q(t_2/t_1)$, which is the same outcome for *HTP* case as in proportional tax credit rule. For *LTP* case, however, $b > t_2/t_1$ generates additional taxation of $\pi_2$ in Country 1, i.e. for the upper threshold $q = 1$, $t_2 < t_1$ implies $b < 1$, and $b$ still has room to increase. MNE achieves the optimal level of transfer pricing manipulation if there is chance to choose $b \leq t_2/t_1$, allowing MNE to equate taxation of $\pi_2$ in both countries. Full repatriation produces $\Pi = (1-t_1)(\pi_1 + \pi_2)$, as in proportional tax credit rule.

Further, we obtain the optimal level of transfer pricing manipulation, with effect of tax restriction, by solving $L_p$ for $p - \bar{p}$; assume the Lagrangian $p - \bar{p} = M(b,q,\lambda)$. Thus,

$$M(b,q,\lambda) = \frac{t_2(1-q) - t_1(1-b)}{\varphi_i z_i} \frac{(p^c - \bar{p})^2}{2} - \lambda(t_1 b - t_2 q)$$

The first order conditions are

$$M_b = \frac{t_1}{\varphi_i z_i}\frac{(p^c-\bar{p})^2}{2} - \lambda t_1 = 0$$

$$M_q = \frac{-t_2}{\varphi_i z_i}\frac{(p^c-\bar{p})^2}{2} + \lambda t_2 = 0$$

and the Kuhn-Tucker condition is $\lambda(t_1 b - t_2 q) = 0$, where $t_1 b \geq t_2 q$, $\lambda \geq 0$. $M_b$ and $M_q$ derive $\lambda = (p^c - \bar{p})^2/2\varphi_i z_i$, so $\lambda > 0$. We arrive at the same solution $b = q(t_2/t_1)$.

---

[25] Assume $r = 2$ in this **Appendix**.

**Appendix D: Effect of repatriation under foreign tax deduction rule**

Differentiating Equation (12) with respect to $b$ provides

$$(p - \bar{p})_b = \frac{t_1(1-t_2)}{\varphi_i z_i} \frac{(p^c - \bar{p})^2}{2}$$

We obtain a distinct "tax differential" term $t_1(1 - t_2)$, which makes $(p - \bar{p})_b > 0$ regardless the tax rates of each country[26]; see that $t_1(1 - t_2)$ is positive for all $t_1 > 0$, $t_2 < 1$ cases[27]. In closer analysis[28], optimal transfer price in Equation (12) implies $p_b > 0$, hence repatriation rate $b$ makes $p - \bar{p}$ to escalate in the direction of a *HTP* and produces different effects for $t_2 > t_1$ and $t_2 < t_1$. For $t_2 > t_1$, increase in $b$ causes an increase in the optimal level of transfer pricing manipulation (decrease in $b$ causes the opposite). For $t_2 < t_1$, otherwise, income shifting incentive is canceled when condition

$$b = \frac{t_1 - t_2}{t_1(1-t_2)}$$

is satisfied[29], and the switch-over spot from *LTP* to *HTP* case is $t_2 = t_1[(1 - b)/(1 - t_1 b)]$, rather than $t_2 = t_1$. From Equation (11) we derive $\Pi_b = -t_1(\pi_2 - t_2\pi_2)$, which is negative for all $t_2 < 1$, and maximization of $\Pi$ occurs when $b = 0$.

**Appendix E: Analysis of the impact of *ad valorem* tariff on imports**

MNE's functions of global net income and the objective function $\phi = \Pi - E(Z_i)$ remain the same as those in previous analysis, but now the profit of subsidiary is $\pi_2 = R_2(s_2) - C_2(s_2 - m) - p(1 + \tau)m$.

First, when home country applies source principle on taxation of foreign profits, optimal level of transfer pricing manipulation is[30]

$$p - \bar{p} = \frac{t_2(1+\tau) - t_1 - \tau}{\varphi_i z_i} \frac{(p^c - \bar{p})^2}{2}$$

Circumstances for *HTP* and *LTP* modify, following the increase of taxation in Country 2: MNE chooses *HTP* if relative tax differential $(t_2 - t_1)/(1 - t_2)$ is higher than tariff $\tau$; otherwise, *LTP* is chosen.

For the proportional tax credit rule, we identify that $t_2 > t_1$ infers $\Pi = (1 - t_1)\pi_1 + (1 - t_2)\pi_2$, so maximization condition is the same as for source-based taxation. For $t_2 < t_1$, optimal $p - \bar{p}$ is

$$p - \bar{p} = \frac{[t_2(1+\tau) - t_1](1-b) - \tau(1-t_1 b)}{\varphi_i z_i} \frac{(p^c - \bar{p})^2}{2}$$

---

[26] This term exhibits the recursive attribute $t_1(1 - t_2) = (1 - t_2)[1 - (1 - t_1)]$.
[27] $t_1 > 0$ denotes the existence of income taxation in home country, and $t_2 < 1$ denotes that income tax rate in Country 2 is less than 100%. This interior case is the prevailing reality for firms and is taken for the analysis of this **Appendix**.
[28] Calling up that $\bar{p}$ is independent, we obtain $(p - \bar{p})_b = p_b > 0$.
[29] This condition is inconsistent for $t_2 > t_1$ because it produces $b < 0$.
[30] Assume $r = 2$ in this **Appendix**.

Comparing it with Equation (8), it shows that repatriation of foreign profits generates a distinct effect than one stated in *Proposition 1*. Differentiating with respect to *b* provides

$$(p - \bar{p})_b = -\frac{t_2(1+\tau) - t_1 - \tau t_1}{\varphi_i z_i} \frac{(p^c - \bar{p})^2}{2}$$

which is different from optimal manipulation under source-based taxation. Repercussion of repatriation rate *b* is weakened in scale $\tau(1 - t_1)$, because $\tau$ reduces net profits of subsidiary, so home taxation of foreign profits is also reduced. Maximization occurs when $b = 0$.

For the limited tax credit rule, inequality constraint $t_1 b \pi_2 \geq t_2 q \pi_2$ remains. Lagrangian function takes the same setup as before, so the maximizing conditions are

$$L_p = t_2(1+\tau)(1-q)m - t_1(1-b)m - \tau(1-t_1 b)m - 2\frac{p-\bar{p}}{(p^c - \bar{p})^2}\varphi_i z_i m - \lambda(t_1 b - t_2 q)(1+\tau)m = 0$$

$$L_b = -t_1 \pi_2 + \lambda t_1 = 0$$

$$L_q = t_2 \pi_2 - \lambda t_2 = 0$$

and the Kuhn-Tucker condition is $\lambda(t_1 b - t_2 q)(1 + \tau)m = 0$, where $t_1 b(1 + \tau)m \geq t_2 q(1 + \tau)m$, $\lambda \geq 0$. $L_p$ derives optimal $p - \bar{p}$, while both $L_b$ and $L_q$ give $\lambda = \pi_2$ (so, $\lambda > 0$). Thus, maximization occurs when $b = q(t_2/t_1)$; this is the same result we obtain for *Proposition 2*, and it derives *Corollary 1* as well[31].

When home country adopts the foreign tax deduction rule, optimal level of transfer pricing manipulation is

$$p - \bar{p} = \frac{[t_2(1+\tau) - \tau](1-t_1 b) - t_1(1-b)}{\varphi_i z_i} \frac{(p^c - \bar{p})^2}{2}$$

and its derivative with respect to *b* provides

$$(p - \bar{p})_b = \frac{t_1(1+\tau)(1-t_2)}{\varphi_i z_i} \frac{(p^c - \bar{p})^2}{2}$$

We have $(p - \bar{p})_b > 0$ for all $t_1 > 0$, $t_2 < 1$; the same conclusion as one analysis of Equation (12). This time, however, a sufficiently high $\tau$ can cause optimal $p - \bar{p}$ to be negative, even for $t_2 > t_1$. Income shifting incentive is neutralized when

$$b = \frac{t_1 - t_2(1+\tau) + \tau}{t_1(1+\tau)(1-t_2)}$$

---

[31] $(p - \bar{p})_b$ gives the same result.

regardless which income tax rate is the highest. Therefore, condition in *Corollary 2* is modified for $\tau > 0$, and it becomes applicable for both *HTP* and *LTP* cases. MNE maximizes global net income when is able to set $b = 0$.